\documentclass[sigconf,nonacm,balance,natbib=true]{acmart}

\usepackage{tikz}
\usetikzlibrary{positioning}
\usetikzlibrary{arrows.meta}
\usetikzlibrary{shapes.symbols}
\usetikzlibrary{fit}
\usepackage{amsmath}
\usepackage{caption}
\usepackage{subcaption}
\usepackage{csquotes}
\usepackage[draft]{minted}
\definecolor{LightGray}{gray}{0.95}

\usepackage{tcolorbox}
\newtcolorbox{lessons}{
    center,
    colframe=gray!90,
    colback=gray!30,
}

\newcommand{\lesson}[1]{
    \begin{lessons}
        \textbf{Lesson}: #1
    \end{lessons}
}

\usepackage{graphicx}

\newcommand{\etcd}{\emph{etcd}}
\newcommand{\metcd}{\emph{mergeable etcd}}
\newcommand{\dismerge}{\emph{dismerge}}
\newcommand{\kubernetes}{\emph{Kubernetes}}

\begin{document}

\title{Mutating \etcd{} Towards Edge Suitability}

\author{Andrew Jeffery}
\orcid{0000-0003-0440-0493}%
\affiliation{
    \institution{University of Cambridge}
    \city{Cambridge}
    \country{United Kingdom}
}
\email{andrew.jeffery@cst.cam.ac.uk}

\author{Heidi Howard}
\orcid{0000-0001-5256-7664}%
\affiliation{
    \institution{Azure Research, Microsoft}
    \city{Cambridge}
    \country{United Kingdom}
}
\email{heidi.howard@microsoft.com}

\author{Richard Mortier}
\orcid{0000-0001-5205-5992}%
\affiliation{
    \institution{University of Cambridge}
    \city{Cambridge}
    \country{United Kingdom}
}
\email{richard.mortier@cst.cam.ac.uk}

\begin{abstract}
In the edge environment servers are no longer being co-located away from clients, instead they are being co-located with clients away from other servers, focusing on reliable and performant operation.
Orchestration platforms, such as \kubernetes{}, are a key system being transitioned to the edge but they remain unsuited to the environment, stemming primarily from their critical key-value stores.
In this work we derive requirements from the edge environment showing that, fundamentally, the design of distributed key-value datastores, such as \etcd{}, is unsuited to meet them.
Using these requirements, we explore the design space for distributed key-value datastores and implement two successive mutations of \etcd{} for different points: \metcd{} and \dismerge{}, trading linearizability for causal consistency based on CRDTs.
\metcd{} retains the linear revision history but encounters inherent shortcomings, whilst \dismerge{} embraces the causal model.
Both stores are local-first, maintaining reliable performance under network partitions and variability, drastically surpassing \etcd{}'s performance, whilst maintaining competitive performance in reliable settings.

The source code this project is available at \\
\url{https://github.com/jeffa5/mergeable-etcd}
\end{abstract}

\maketitle

\section{Introduction}\label{sec:introduction}

More compute resources are becoming available near the edge of the network leading to an increasing interest in deploying services there.
These services can perform aggregation closer to the edge, reducing the volume of data to be sent to the cloud as well as offering clients more local operations~\cite{cfn}.
They can typically be deployed in mini data centers~\cite{minidcs} --- small, mostly ISP operated, compute sites.
With each site being geographically distributed, networks between edge sites can have higher latency than intra-datacenter communication coupled with increased likelihood of network partitions.
This is further exacerbated by resource limitations at each site, requiring efficient use of those resources.

Resource aggregation is critical to this environment, exploiting the numerous but geodistributed resources each site offers.
Aggregating sites into larger clusters enables running larger jobs with higher availability, capitalising on the deployed resources and the periodicity of demand.
A single large cluster also eases management and operation of the services, offering them higher availability across sites through efficient orchestration.

\kubernetes{}~\cite{kubernetes} is a \emph{container} orchestration platform based on Google's Borg~\cite{borg} system.
\kubernetes{} is used by a majority of the top 500 companies in the world~\cite{trustk8s}, managing deployments of services over thousands of nodes, handling failures automatically~\cite{openai}.
In large data centers it leverages the low latency networks, having a centralised control-plane and datastore, \etcd{}~\cite{etcd}, for coordinating the various actions.
Despite its prevalence in data centers, there is growing interest in deploying it to mini data centers close to the edge with specific projects targeting this use case~\cite{k3s, kubeedge}.

\etcd{} is a distributed, but logically centralised, key-value store.
It is widely used for cloud applications, including \kubernetes{}, \emph{Rook}, \emph{CoreDNS} and \emph{M3}~\cite{etcd}.
Due to its critical place in these systems it is a key factor for them being suitable to deploy to the edge.
In fact, \etcd{} has already been shown to have scalability limitations under best-case scenarios~\cite{rearchk8s}, which would only be exacerbated at the network edge with its higher latency cross-site links.
As \etcd{} is critical to cloud applications' operation, they are also bounded by \etcd{}'s ability to tolerate higher latencies and network faults, impacting scalability and reliability~\cite{k8sissues, rookissues, m3issues, corednsissues}.

In the process of analysing and deriving requirements from the edge environment we present the design and implementation of two successive adaptations to \etcd{}: \metcd{} and \dismerge{} trading linearizability~\cite{herlihy1990linearizability, viotti2016consistency} for causal consistency~\cite{lamport2019time, mahajan2011consistency, viotti2016consistency} with Conflict-free Replicated DataTypes (CRDTs)~\cite{Letia2010consistency, Shapiro2011CRDTs}.
These target the edge environment with limited heterogeneous resources whilst maintaining as close semblance to \etcd{} as possible to minimise programming model differences and thus respective changes in the systems built around \etcd{}.
They explore two different points in the design space, \metcd{} focusing on maintaining compatibility with \etcd{} and its linear history, and \dismerge{} exploring the impacts of changes of exposing the causal history explicitly.
From these design choices, we show that both datastores maintain consistent performance under network partitions and variability, surpassing \etcd{}'s performance, whilst also remaining competitive in reliable settings at the edge.
Our contributions in this paper are as follows:
\begin{enumerate}
    \item We analyse the requirements for edge focused distributed key-value stores, Section~\ref{sec:edgekvs}.
    \item We outline design trade-offs to cater for these requirements, Section~\ref{sec:design}.
    \item We present the implementation of the two datastores exploring different parts of this design space, Section~\ref{sec:implementation}.
    \item We evaluate the systems highlighting \metcd{}'s and \dismerge{}'s ability to operate with consistent performance under larger cluster sizes and added latency, Section~\ref{sec:eval}.
    \item We discuss the implications of the changes applied on broader systems, particularly \kubernetes{}, Section~\ref{sec:discussion}.
\end{enumerate}

\section{Background and motivation}\label{sec:edgekvs}

\subsection{\etcd{}}\label{sec:etcd}

\begin{figure}
    \centering
    \includegraphics[width=\linewidth]{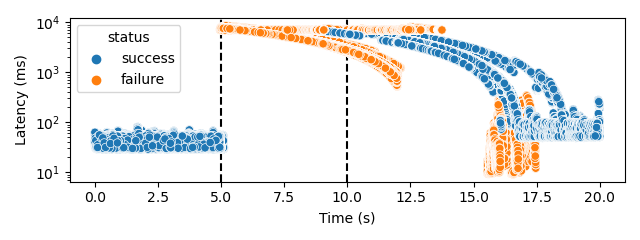}
    \caption{Impact of a network partition on a 3 node \etcd{} cluster. See Section~\ref{sec:eval-setup} for more details.}\label{fig:etcd-fault-tolerance}
    \Description{Scatter plot showing the latency of individual requests over time. It initially starts around 30ms before the partition, then jumps to timeouts during the partition before coming back down after healing, with a delay as it works through the queue. Finally it settles at a higher latency, closer to 100ms.}
\end{figure}

\etcd{} is a Raft-based~\cite{raft} linearizable distributed key-value store, requiring majority quorums.
It exposes a straightforward API with the ability to get ranges of values, write values and delete ranges as well as being able to do these within transactions, all over a single flat key-space.
Another aspect of \etcd{} is its ability for clients to \emph{watch} values and get updates pushed to them directly, through \emph{watch streams}, without the need to poll.
Due to its use of linearizability \etcd{} can struggle to perform adequately at scale~\cite{rearchk8s}, this is fundamentally limited by the fault-tolerance model it adopts~\cite{consensusinpartialsynchrony}.
Since only a leader can process write requests, or linearizable reads, client requests must either target the leader directly or be forwarded, adding extra latency out of the control of the client.
Due to its fault-tolerance model \etcd{} is unable to process requests without communicating with a majority of nodes (Figure~\ref{fig:etcd-fault-tolerance}), leaving partitioned sites unable to adapt.
\etcd{} can also exhibit subtle failure conditions under misbehaving networks~\cite{raftbehaviour}.

\etcd{} makes the following guarantees about its Key-Value API~\cite{etcdkvguarantees}:
\begin{description}
    \item[Atomicity]  Operations complete entirely or not at all.
    \item[Durability] Completed operations are durable and a read operation never returns data that is not durable.
    \item[Consistency] Operations are linearizable.
    \item[Completeness of watches] Watch events never observe partial events for a single operation.
    \item[Global revision] Each mutating request is assigned a strictly monotonically increasing revision number.
\end{description}

\subsection{Edge environment}

We focus on miniature data centers and compute at the network edge.
These sites are resource constrained in multiple dimensions: CPU, memory, and networking.
Near-edge compute sites are typically small but larger in number to provide closer operation to the user.
This large scale places emphasis on avoiding overheads from cross-site communication which can be costly but also unreliable in latency, bandwidth and consistency due to competing with user traffic.

Applications running at the edge and serving user traffic want low latency operation, to be able to handle a dynamic environment, avoid cross-site dependencies and be able to progress independently of other sites.

\subsection{Deriving requirements}

From the characteristics of the edge environment and the expectations of applications relying on datastores such as \etcd{}, we derive the following requirements for datastores deployed at the edge:

\paragraph{Site-local reads}
To serve applications with low latency and avoid cross-site communication reads need to be site-local.
This can be viewed similarly to a content-delivery network~\cite{cdn} which has content cached at the edge to reduce latency of operations.
Implied by site-local reads, each node needs to maintain all historical data for each key locally.
This limits the overall quantity of data that can be stored but is key in enabling site-local reads with history.

\paragraph{Site-local writes}
Further to site-local reads we also want a system that supports site-local writes.
This ensures that the system can operate even when network connectivity is impaired.

\paragraph{Performance}
Since edge applications need to be performant for user expectations as well as supporting lots of work at the edge we require the datastore to be performant.

\paragraph{Resource efficiency}
In addition to performance, we want our application to be efficient in its storage, using a small overhead compared to the raw data storage requirement.

Of these requirements, \etcd{} is only able to fulfil site-local reads when serializable reads are used, which is uncommon in our experience.
Site-local writes are never possible in \etcd{} clusters of more than one node.
Performance will be covered more in the evaluation section (Section~\ref{sec:eval}), but its architecture is targeted towards cloud data center deployments.
Due to this targeting, it is also not the most resource efficient, ideally running on large multi-core machines.

\subsection{Application deployment}\label{sec:deployment}

\begin{figure*}
    \centering
    \begin{subfigure}[b]{0.24\linewidth}
        \centering
        \begin{tikzpicture}
            \node[circle, fill=red!70] (master1) [] {};
            \node[circle, fill=red!70] (master2) [right=0.5cm of master1] {};
            \node[circle, fill=red!70] (master3) [right=0.5cm of master2] {};

            \node[circle, fill=cyan] (worker1) [below=0.5cm of master1] {};
            \node[circle, fill=cyan] (worker2) [right=0.5cm of worker1] {};
            \node[circle, fill=cyan] (worker3) [right=0.5cm of worker2] {};

            \node[circle, fill=cyan] (worker4) [below=0.5cm of worker1] {};
            \node[circle, fill=cyan] (worker5) [right=0.5cm of worker4] {};
            \node[circle, fill=cyan] (worker6) [right=0.5cm of worker5] {};

            \node[draw, cloud, aspect=1.2, scale=0.8, fit=(master1)(master2)(master3)(worker1)(worker2)(worker3)(worker4)(worker5)(worker6)] (dc1) {};

            \draw[Latex-Latex] (master1) -- (master2);
            \draw[Latex-Latex] (master2) -- (master3);

            \draw[Latex-Latex] (master1) -- (worker1);
            \draw[Latex-Latex] (master2) -- (worker2);
            \draw[Latex-Latex] (master3) -- (worker3);
            \draw[Latex-Latex] (master2) -- (worker4);
            \draw[Latex-Latex] (master3) -- (worker5);
            \draw[Latex-Latex] (master2) -- (worker6);
        \end{tikzpicture}
        \caption{All-cloud.}\label{fig:traditional-deployment}
        \Description{A diagram showing all worker nodes are able to connect to the control plane nodes without leaving the site.}
    \end{subfigure}
    \hfill
    \begin{subfigure}[b]{0.24\linewidth}
        \centering
        \begin{tikzpicture}
            \node[circle, fill=red!70] (master1) [] {};
            \node[circle, fill=red!70] (master2) [right=0.5cm of master1] {};
            \node[circle, fill=red!70] (master3) [right=0.5cm of master2] {};

            \node[circle, fill=cyan] (worker1) [below=0.5cm of master1] {};
            \node[circle, fill=cyan] (worker2) [right=0.5cm of worker1] {};
            \node[circle, fill=cyan] (worker3) [right=0.5cm of worker2] {};

            \node[circle, fill=cyan] (worker4) [below=0.5cm of worker1] {};
            \node[circle, fill=cyan] (worker5) [right=0.5cm of worker4] {};
            \node[circle, fill=cyan] (worker6) [right=0.5cm of worker5] {};

            \node[draw, rounded corners, fit=(master1)(worker1)(worker4)] (dc1) {};
            \node[draw, rounded corners, fit=(master2)(worker2)(worker5)] (dc2) {};
            \node[draw, rounded corners, fit=(master3)(worker3)(worker6)] (dc3) {};

            \draw[Latex-Latex] (worker1) -- (master1);
            \draw[Latex-Latex] (worker2) -- (master2);
            \draw[Latex-Latex] (worker3) -- (master3);
            \draw[Latex-Latex] (worker4) edge [out=110,in=250] (master1);
            \draw[Latex-Latex] (worker5) edge [out=110,in=250] (master2);
            \draw[Latex-Latex] (worker6) edge [out=110,in=250] (master3);
        \end{tikzpicture}
        \caption{k3s: single-site.}\label{}
        \Description{A diagram showing multiple sites, each only using local communication.}
    \end{subfigure}
    \hfill
    \begin{subfigure}[b]{0.24\linewidth}
        \centering
        \begin{tikzpicture}
            \node[circle, fill=red!70] (master1) [] {};
            \node[circle, fill=red!70] (master2) [right=0.5cm of master1] {};
            \node[circle, fill=red!70] (master3) [right=0.5cm of master2] {};

            \node[circle, fill=cyan] (worker1) [above=0.5cm of master1] {};
            \node[circle, fill=cyan] (worker2) [right=0.5cm of worker1] {};
            \node[circle, fill=cyan] (worker3) [right=0.5cm of worker2] {};

            \node[circle, fill=cyan] (worker4) [below=0.5cm of master1] {};
            \node[circle, fill=cyan] (worker5) [right=0.5cm of worker4] {};
            \node[circle, fill=cyan] (worker6) [right=0.5cm of worker5] {};

            \node[draw, rounded corners, fit=(master1)(worker1)] (dc1) {};
            \node[draw, rounded corners, fit=(master2)(worker2)] (dc2) {};
            \node[draw, rounded corners, fit=(master3)(worker3)] (dc3) {};
            \node[draw, rounded corners, fit=(worker4)] (dc4) {};
            \node[draw, rounded corners, fit=(worker5)] (dc5) {};
            \node[draw, rounded corners, fit=(worker6)] (dc6) {};

            \draw[Latex-Latex] (master1) -- (master2);
            \draw[Latex-Latex] (master2) -- (master3);

            \draw[Latex-Latex] (worker1) -- (master1);
            \draw[Latex-Latex] (worker2) -- (master2);
            \draw[Latex-Latex] (worker3) -- (master3);
            \draw[Latex-Latex] (worker4) -- (master1);
            \draw[Latex-Latex] (worker5) -- (master2);
            \draw[Latex-Latex] (worker6) -- (master3);
        \end{tikzpicture}
        \caption{Vanilla: multi-site.}\label{}
        \Description{A diagram showing multiple sites, some are able to operate locally, others must leave their site to reach a control plane node. The control-plane nodes are collaborative.}
    \end{subfigure}
    \hfill
    \begin{subfigure}[b]{0.24\linewidth}
        \centering
        \begin{tikzpicture}
            \node[circle, fill=red!70] (master1) [] {};
            \node[circle, fill=red!70] (master2) [right=0.5cm of master1] {};
            \node[circle, fill=red!70] (master3) [right=0.5cm of master2] {};

            \node[circle, fill=cyan] (worker1) [below=1cm of master1] {};
            \node[circle, fill=cyan] (worker2) [right=0.5cm of worker1] {};
            \node[circle, fill=cyan] (worker3) [right=0.5cm of worker2] {};

            \node[circle, fill=cyan] (worker4) [below=0.5cm of worker1] {};
            \node[circle, fill=cyan] (worker5) [right=0.5cm of worker4] {};
            \node[circle, fill=cyan] (worker6) [right=0.5cm of worker5] {};

            \node[draw, cloud, aspect=3, scale=0.8, fit=(master1)(master2)(master3)] (dc1) {};
            \node[draw, rounded corners, fit=(worker1)] (dc4) {};
            \node[draw, rounded corners, fit=(worker2)] (dc5) {};
            \node[draw, rounded corners, fit=(worker3)] (dc6) {};
            \node[draw, rounded corners, fit=(worker4)] (dc7) {};
            \node[draw, rounded corners, fit=(worker5)] (dc8) {};
            \node[draw, rounded corners, fit=(worker6)] (dc9) {};

            \draw[Latex-Latex] (master1) -- (master2);
            \draw[Latex-Latex] (master2) -- (master3);

            \draw[Latex-Latex] (master1) -- (worker1);
            \draw[Latex-Latex] (master2) -- (worker2);
            \draw[Latex-Latex] (master3) -- (worker3);
            \draw[Latex-Latex] (master2) -- (worker4);
            \draw[Latex-Latex] (master3) -- (worker5);
            \draw[Latex-Latex] (master2) -- (worker6);
        \end{tikzpicture}
        \caption{KubeEdge: cloud-centric.}\label{}
        \Description{A diagram showing all worker nodes in separate edge sites and all control-plane nodes operating in the cloud. Worker nodes must reach the cloud in order to perform actions.}
    \end{subfigure}
    \begin{subfigure}[b]{\linewidth}
        \centering
        \begin{tikzpicture}
            \node[circle, fill=red!70] (red) [] {};
            \node[] (redtext) [right=0.1cm of red] {Control plane node};
            \node[circle, fill=cyan] (blue) [right=0.5cm of redtext] {};
            \node[] (bluetext) [right=0.1cm of blue] {Worker node};
        \end{tikzpicture}
    \end{subfigure}
    \caption{Edge application deployment strategies. Boxes indicate edge sites, arrows indicate potential connections.}\label{fig:edge-deployments}
\end{figure*}
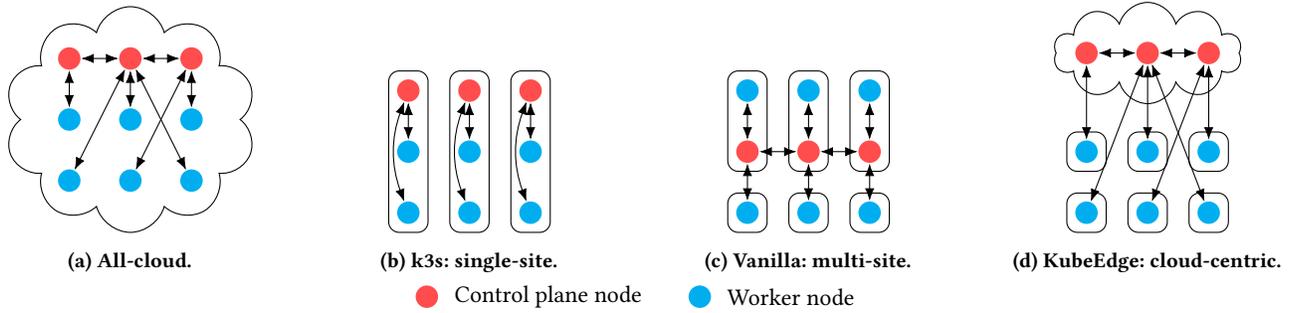

Given deployments of \etcd{} to the edge, we observe three main strategies based off \kubernetes{}: single-site (K3s)~\cite{k3s}, cross-site (vanilla)~\cite{kubernetes}, and cloud-centric (KubeEdge)~\cite{kubeedge}.
Figure~\ref{fig:edge-deployments} shows the layout of these and Table~\ref{tab:edge-deployments} highlights the requirements they satisfy from the point of view of a single edge site, assuming \etcd{} would be deployed at each control plane node.
Blast radius considers what would be impacted if a site with control-plane node gets disconnected from everything else.

Datastores based on eventual consistency, such as Cassandra~\cite{cassandra}, can be deployed in equivalent configurations but still do not satisfy the requirements.
Since data is partitioned across nodes, each node does not store all data, violating the site-local reads requirement, writes are also not guaranteed to be served locally, depending on replication requirements.

\begin{table*}
    \caption{Comparison of requirements met by \etcd{} deployed with deployment strategies from Figure~\ref{fig:edge-deployments}. }\label{tab:edge-deployments}
    \begin{tabular}{llllll}
        \toprule
        Case          & Site-local reads & Site-local writes & Efficiency       & Management       & Blast radius   \\
        \midrule
        All-cloud     & Yes              & Yes               & Great            & Single cluster   & Single site    \\
        Single-site   & Yes              & Yes               & Wasted resources & Lots of clusters & Single site    \\
        Multi-site    & No               & No                & Great            & Single cluster   & Multiple sites \\
        Cloud-centric & No               & No                & Bandwidth cost   & Single cluster   & All edge sites \\
        \bottomrule
    \end{tabular}
\end{table*}

Running small clusters of datastores such as \etcd{} at the center of large systems such as \kubernetes{} leaves the large systems vulnerable to broader faults, particularly at the edge.
As these systems become distributed across data centers for fault-tolerance, or edge sites for locality, they may retain access to only one datastore node.
When this datastore node becomes unable to process requests, due to failure, all attached clients are unable to perform their actions.
This creates a very large blast radius for the core distributed key-value store, commonly relying on majority replication with a cluster size of 3 or 5.

\section{Design space}\label{sec:design}

Table~\ref{tab:store-properties} highlights the key differences in the datastores presented.
This focuses around four primary points in the design space: consistency of data, how history is addressed, durability of data, and how values are represented.
In this section we explore the choices each datastore makes within these parameters.

\begin{table*}
    \caption{Comparison of properties of the datastores.}\label{tab:store-properties}
    \begin{tabular}{llrlll}
        \toprule
        Store       & Consistency  & Fault tolerance & History addressing & Durability        & Values           \\
        \midrule
        \etcd{}     & Linearizable & \(2f+1\)        & Integer counter    & Majority of nodes & Bytes            \\
        \metcd{}    & Causal       & \(f+1\)         & Integer counter    & Single node       & Operator-defined \\
        \dismerge{} & Causal       & \(f+1\)         & Hash graph heads   & User dependent    & Operator-defined \\
        \bottomrule
    \end{tabular}
\end{table*}

\subsection{Consistency and fault tolerance}

\lesson{Strong consistency is an availability and scalability bottleneck.}

\etcd{} uses strong consistency, particularly linearizability, to replicate values between stores.
This means that, for \(f\) node failures, it requires \(2f+1\) to be in the cluster.
In cloud environments, \etcd{} can make assumptions of node homogeneity, for both node sizes and network links.
However, near the edge these assumptions, particularly those of the network links, may not hold.
This impacts the scalability of the cluster, and ultimately the availability it can provide.
Therefore, the heterogeneous nature of the edge leads to the imbalance of fault tolerance across sites explored in Section~\ref{sec:deployment}.
Since \etcd{} is the critical core of many systems, it is notable that this limitation of fault-tolerance directly impacts systems considerably bigger than itself.

A weaker variant of using linearizability, which enables stronger availability, is causal consistency.
This can be easily implemented with CRDTs.
This model enables the data viewed at different nodes of a system to differ, with the guarantee that it will converge in the steady-state.
In practice, this enables pushing replication of updates between nodes from happening eagerly to happening lazily.
This decouples nodes, enabling them to tolerate more heterogeneous network links, including handling updates whilst experiencing complete partitions from the cluster.
To tolerate \(f\) node failures these systems require only \(f+1\) nodes in the cluster.
This decoupling also enables these clusters to scale better, being able to match the deployment scale of edge sites.
This makes the applications built on these systems able to be more performant and reliable.

\subsubsection{Mutable histories}

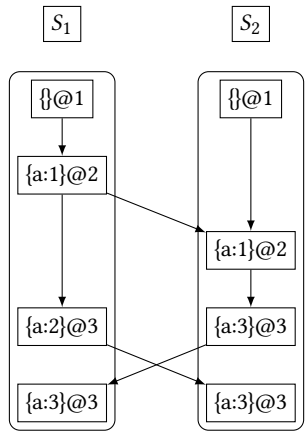
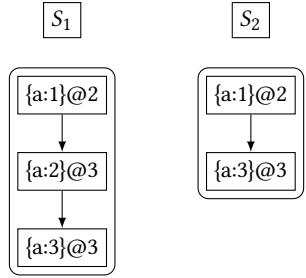
\begin{figure}
    \begin{subfigure}[b]{\linewidth}
        \centering
        \begin{tikzpicture}
            \node[draw] (s1) {\(S_1\)};
            \node[draw] (s2) [right=2cm of s1] {\(S_2\)};
            \node[draw] (a) [below=0.5cm of s1] {\{\}@1};
            \node[draw] (b) [below=0.5cm of s2] {\{\}@1};
            \node[draw] (c) [below=0.5cm of a] {\{a:1\}@2};
            \node[draw] (f) [below=1.5cm of b] {\{a:1\}@2};
            \node[draw] (g) [below=1.5cm of c] {\{a:2\}@3};
            \node[draw] (h) [below=0.5cm of f] {\{a:3\}@3};
            \node[draw] (i) [below=0.5cm of g] {\{a:3\}@3};
            \node[draw] (j) [below=0.5cm of h] {\{a:3\}@3};

            \node[draw, rounded corners, fit=(a)(c)(g)(i)] (s1group) {};
            \node[draw, rounded corners, fit=(b)(f)(h)(j)] (s2group) {};

            \draw[-latex] (a) -- (c);
            \draw[-latex] (c) -- (g);
            \draw[-latex] (b) -- (f);
            \draw[-latex] (f) -- (h);
            \draw[-latex] (c) -- (f);
            \draw[-latex] (g) -- (j);
            \draw[-latex] (h) -- (i);
        \end{tikzpicture}
        \caption{Sequence of updates to two \metcd{} datastores. History is mutable (revision 3 on \(S_1\)).}\label{fig:updates}
        \Description{A diagram showing two servers performing changes and synchronizing. Server 1 writes a:1, which gets revision 2. This is then synchronized to server 2. Server 1 then writes a:1, which gets revision 3 whilst Server 2 concurrently writes a:3 which also gets revision 3. Both servers synchronize their changes and the result is Server 2's write.}
    \end{subfigure}

    \begin{subfigure}[b]{\linewidth}
        \centering
        \begin{tikzpicture}
            \node[draw] (s1) {\(S_1\)};
            \node[draw] (s2) [right=2cm of s1] {\(S_2\)};
            \node[draw] (a) [below=0.5cm of s1] {\{a:1\}@2};
            \node[draw] (b) [below=0.5cm of a] {\{a:2\}@3};
            \node[draw] (e) [below=0.5cm of b] {\{a:3\}@3};
            \node[draw] (c) [below=0.5cm of s2] {\{a:1\}@2};
            \node[draw] (d) [below=0.5cm of c] {\{a:3\}@3};

            \node[draw, rounded corners, fit=(a)(b)(e)] (s1group) {};
            \node[draw, rounded corners, fit=(c)(d)] (s2group) {};

            \draw[-latex] (a) -- (b);
            \draw[-latex] (b) -- (e);
            \draw[-latex] (c) -- (d);
        \end{tikzpicture}
        \caption{Sequence of corresponding watch updates.}\label{fig:watches}
        \Description{A diagram showing the same two servers' watch event stream. Server 1's stream observes a:1 at revision 2, a:2 at revision 3 and then a:3 at revision 3. Server 2's stream observers a:1 at revision 2 followed by a:3 at revision 3.}
    \end{subfigure}
    \caption{Updates and watches at \metcd{}. Notation in the form \(\{key:value\}@revision\).}%
\end{figure}

One challenge in adapting the data model of \etcd{} to work with causal consistency is that the previously immutable history becomes mutable.
Figure~\ref{fig:updates} shows the process of two peers synchronizing whilst having writes from separate clients.
The first write is to \(S_1\) which synchronizes with \(S_2\) without it having concurrent writes, so they both remain consistent.
However, both nodes then receive concurrent writes to the same key, \(a\).
This means that they will both use the same revision for this update, \(3\), but have different values for the key.
When they next synchronize this value needs to be made consistent across the replicas and in this case the value from \(S_2\) wins over the value from \(S_1\).
If the client who last wrote to \(S_1\) retrieves the value for \(a\) again, it will see the updated value \(3\) at the same revision.
This mutable history is a consequence of the causal consistency coupled with \etcd{}'s global revision counter.

Due to lazy synchronizations, datastores can have an imbalance of updates made to them.
If the same key is altered on different nodes concurrently then upon a merge the one with the higher revision may dominate the other.
This can even be due to updates on other keys in the store, artificially progressing the revision counter before the same key is then updated.
This dominating behaviour is worst when synchronization is infrequent, particularly likely in times of failures such as network partitions.
\metcd{} is more vulnerable to this behaviour than \dismerge{} due to the way that they address changes.

\subsubsection{Watching values}\label{sec:mergeable-watches}

When a client requests a stream of watch events from a server it is guaranteed to observe complete changes, knowing the history is immutable.
Since the history can change in \metcd{}, two watch streams (connected to different servers) may observe different values at the same revision, breaking this guarantee.
When the two servers synchronize they will have a consistent view of the values but the clients may not be updated with the result of this conflict-resolution.
When synchronizing the servers can send watch events for values if the revision is newer, or even the same as that last sent as long as the incoming value is the \emph{winner}.
For example, in Figure~\ref{fig:watches} the server \(S_1\) would send the new update for revision 3 whilst server \(S_2\) does not need to as it has already sent that value.
The first client will have a local conflict and so should forget its past value and accept the newer one, whilst the latter client retains the original value.

\subsection{Addressing history}

\lesson{Linear histories prevent all changes being addressed under causal consistency.}

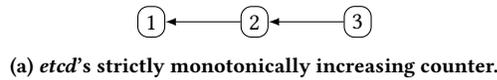
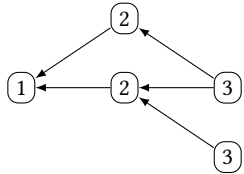
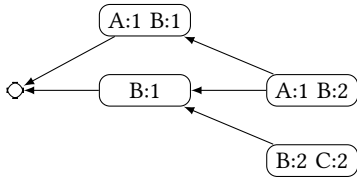
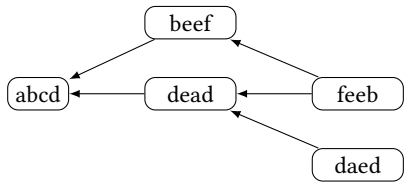
\begin{figure}
    \centering
    \begin{subfigure}[b]{\linewidth}
        \centering
        \begin{tikzpicture}
            \node[draw, rounded corners] (1) {1};
            \node[draw, rounded corners] (2) [right=of 1] {2};
            \node[draw, rounded corners] (3) [right=of 2] {3};

            \draw[-latex] (2) -- (1);
            \draw[-latex] (3) -- (2);
        \end{tikzpicture}
        \caption{\etcd{}'s strictly monotonically increasing counter.}\label{fig:etcdhistory}
        \Description{A graph showing a linear chain of revisions: 1, 2, 3.}
    \end{subfigure}

    \begin{subfigure}[b]{\linewidth}
        \centering
        \begin{tikzpicture}
            \node[draw, rounded corners] (1) {1};
            \node[draw, rounded corners] (2a) [right=of 1] {2};
            \node[draw, rounded corners] (2b) [above=0.5cm of 2a] {2};
            \node[draw, rounded corners] (3a) [right=of 2a] {3};
            \node[draw, rounded corners] (3b) [below=0.5cm of 3a] {3};

            \draw[-latex] (2a) -- (1);
            \draw[-latex] (2b) -- (1);
            \draw[-latex] (3a) -- (2a);
            \draw[-latex] (3a) -- (2b);
            \draw[-latex] (3b) -- (2a);
        \end{tikzpicture}
        \caption{\metcd{}'s counter with concurrent edits.}\label{fig:metcdhistory}
        \Description{A graph showing a branching chain of revisions with duplicates.}
    \end{subfigure}

    \begin{subfigure}[b]{\linewidth}
        \centering
        \begin{tikzpicture}
            \node[draw, rounded corners] (1) {};
            \node[draw, rounded corners, text width=1cm, align=center] (2a) [right=of 1] {B:1};
            \node[draw, rounded corners, text width=1cm, align=center] (2b) [above=0.5cm of 2a] {A:1 B:1};
            \node[draw, rounded corners, text width=1cm, align=center] (3a) [right=of 2a] {A:1 B:2};
            \node[draw, rounded corners, text width=1cm, align=center] (3b) [below=0.5cm of 3a] {B:2 C:2};

            \draw[-latex] (2a) -- (1);
            \draw[-latex] (2b) -- (1);
            \draw[-latex] (3a) -- (2a);
            \draw[-latex] (3a) -- (2b);
            \draw[-latex] (3b) -- (2a);
        \end{tikzpicture}
        \caption{Vector clock-based change addressing.}\label{fig:vectorhistory}
        \Description{A diagram showing the same branching chain of revisions but using vector clocks for two actors A and B.}
    \end{subfigure}

    \begin{subfigure}[b]{\linewidth}
        \centering
        \begin{tikzpicture}
            \node[draw, rounded corners] (1) {abcd};
            \node[draw, rounded corners, text width=1cm, align=center] (2a) [right=of 1] {dead};
            \node[draw, rounded corners, text width=1cm, align=center] (2b) [above=0.5cm of 2a] {beef};
            \node[draw, rounded corners, text width=1cm, align=center] (3a) [right=of 2a] {feeb};
            \node[draw, rounded corners, text width=1cm, align=center] (3b) [below=0.5cm of 3a] {daed};

            \draw[-latex] (2a) -- (1);
            \draw[-latex] (2b) -- (1);
            \draw[-latex] (3a) -- (2a);
            \draw[-latex] (3a) -- (2b);
            \draw[-latex] (3b) -- (2a);
        \end{tikzpicture}
        \caption{\dismerge{}'s hash-based change addressing. Each node represents the hex-encoded hash.}\label{fig:dismergehistory}
        \Description{A diagram showing the same branching chain of revisions but using hashes.}
    \end{subfigure}

    \caption{Revision representations visualised.}\label{fig:dag}
\end{figure}

\etcd{} maintains the history of all values, making them addressable with an integer counter, Figure~\ref{fig:etcdhistory}.
This provides users with a unique handle for changes which they can use to look back in time, or resume watch streams from a known last position.
This counter is suitable under linearizability as there can only be one update for each revision.
With causal consistency, this breaks down because changes can be made to multiple nodes in parallel, thus they may get the same revision assigned.
When the nodes synchronize, the updates will effectively conflict in the history space, breaking the expectation that the revision counter is a unique handle, Figure~\ref{fig:metcdhistory}.
Additionally, updates synchronized from nodes can appear in the past.
This poses challenges for sending updates over watch streams as the clients expect to already have observed the latest version, and so should not be sent an update for a past revision.
However, due to the nature of the update clients may care about it and wish to update the value after merging the representations, this is not possible using the single counter revisions.

Instead, when multiple nodes are accepting updates, we can use vector clocks to tag the updates, forming a directed acyclic graph (DAG) of changes, Figure~\ref{fig:vectorhistory}.
This has the advantage that now every update has a unique identifier but the downside of the clocks growing, without removal.
The clocks will grow linearly in size \(O(n)\) for \(n\) nodes in the cluster, which is large near the edge.
These clocks would be included in every request to identify the current \emph{revision} for clients.
Rather than incur the overhead of sending these clocks over the network, we can view the updates as a hash DAG, similar to that of Git~\cite{git}, Figure~\ref{fig:dismergehistory}.
Each update is uniquely represented by a single hash, which encompasses the operations in the update itself along with the hashes of its ancestors, scaling with \(O(1)\) independently of the size of the cluster.
This equates to every change being a \enquote{merge commit} of the frontier of the DAG.\@
Since changes are now uniquely and efficiently addressed clients can always view the history at the point in time of each individual hash, or provide a group of hashes to observe the data at a point where multiple changes are simultaneously visible.

Clients can obtain the current set of frontier hashes for a node.
However, unlike the revision counter from \etcd{}, the set of frontier hashes is not guessable or predictable for clients.
However, the revision field is typically used for addressing the \emph{observed} history of the datastore, particularly during watch streams.
When clients request watch updates for keys, they maintain a record of the last revision they encountered from an update.
When they restart they can use this as an opaque identifier to the datastore as a placeholder to pick up from where they last observed.
Since the revision counter is treated as opaque, the frontier hashes can be used similarly.

\subsection{Durability}

\lesson{Lack of individual change addressing leads to difficult durability management.}

When \etcd{} replicates changes to other nodes, obtaining consensus over them, the changes are made durable at each node before they acknowledge it.
This ensures that, even in the event that the entire cluster restarts simultaneously, the change will still be accessible.
When replication is lazy, as with causal consistency, the change is only made durable on the node processing the change before responding to the client.
Upon replicating the change to other nodes it becomes durable on them, however, since this is a background process the client has no information about which nodes have received a given change.

As we have seen in the previous section, a revision counter prevents individual changes being addressed, posing an issue for detecting what nodes have made it durable.
Importantly, a single revision counter means that clients must assume the change only ever has durability at the node it was performed at.
However, using hashes for changes, and making them uniquely addressable, we regain the ability to query nodes for their durable changes.
The information of what changes each node has can be included in the synchronization protocol such that a single node will be able to inform a client of the replication status of a change made there.
Clients can then use this information to wait for a particular replication threshold for their changes to suit them.

\subsection{Value representation}

\lesson{Introspecting values at the datastore can provide semantic updates.}

\begin{listing}
    \centering
    \begin{minted}[frame=lines, framesep=2mm, bgcolor=LightGray, fontsize=\small]{rust}
#[derive(Reconcile, Hydrate, Serialize, Deserialize)]
struct Deployment {
  image: String,
  replicas: u32,
}
    \end{minted}
    \caption{Example of using typed values. Based on a \kubernetes{} \emph{Deployment} resource.}\label{lst:automergeable}
\end{listing}

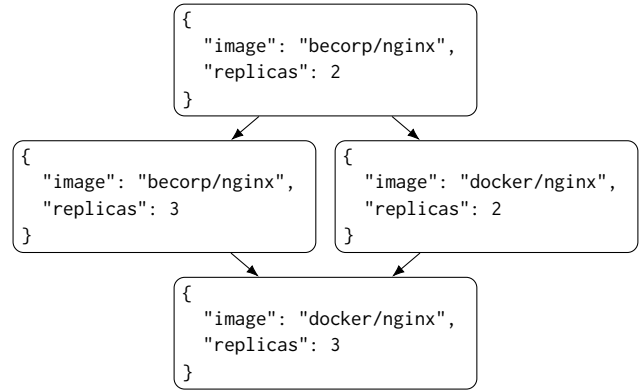
\begin{figure}
    \centering
    \begin{tikzpicture}
        \node[draw, rounded corners] (root) {
            \begin{minipage}{0.45\linewidth}
                \begin{minted}[fontsize=\small]{json}
{
  "image": "becorp/nginx",
  "replicas": 2
}
                    \end{minted}
            \end{minipage}
        };
        \node[draw, rounded corners] (left) [below left=1em and -6em of root] {
            \begin{minipage}{0.45\linewidth}
                \begin{minted}[fontsize=\small]{json}
{
  "image": "becorp/nginx",
  "replicas": 3
}
                    \end{minted}
            \end{minipage}
        };
        \node[draw, rounded corners] (right) [below right=1em and -6em of root] {
            \begin{minipage}{0.45\linewidth}
                \begin{minted}[fontsize=\small]{json}
{
  "image": "docker/nginx",
  "replicas": 2
}
                    \end{minted}
            \end{minipage}
        };
        \node[draw, rounded corners] (bottom) [below right=1em and -6em of left] {
            \begin{minipage}{0.45\linewidth}
                \begin{minted}[fontsize=\small]{json}
{
  "image": "docker/nginx",
  "replicas": 3
}
                    \end{minted}
            \end{minipage}
        };

        \draw[-Latex] (root) -- (left);
        \draw[-Latex] (root) -- (right);
        \draw[-Latex] (left) -- (bottom);
        \draw[-Latex] (right) -- (bottom);
    \end{tikzpicture}
    \caption{Example of concurrently modifying two values, based on the datatype from Listing~\ref{lst:automergeable}.}\label{fig:datatypes}
    \Description{A diagram showing concurrent mutations to a Deployment struct being merged correctly.}
\end{figure}

Treating the values as opaque bytes, as \etcd{} does, can make for efficient, and application agnostic, handling of requests.
If \etcd{} were to support structured values, such as JSON, it would still be going through consensus on the individual updates, despite them potentially being to distinct parts of the datatype.
By enabling concurrent writes with \metcd{} and \dismerge{} using raw bytes for values, the conflict-resolution is very coarse-grained, being at the level of entire values.
Supporting introspection of the value, based on a datatype, natively enables the datastore to be able to provide more fine-grained conflict-resolution, such as allowing concurrent mutations to different parts of the datatype.
For instance, for orchestration workloads we may have two controllers operating concurrently that perform separate jobs.
One is responsible for updating the image to point to the correct location, the other is an autoscaler, responsible for ensuring enough instances of the application are available to handle the demand.
In \etcd{}, these updates must happen one before the other, requiring the second to re-apply the update locally before sending to the datastore again, effectively being last-writer wins.
With \metcd{} and \dismerge{} though, the updates do not need to be strictly ordered, they will merge together when both changes are present at a datastore node.
Figure~\ref{fig:datatypes} highlights this difference based on the datatype in Listing~\ref{lst:automergeable}.

\section{Implementation}\label{sec:implementation}

\metcd{} and \dismerge{} share a similar architecture, serving an \etcd{}-like API but based around a CRDT document to enable decentralized operation.
They are both implemented in Rust using the Automerge CRDT library at the core.
The Automerge CRDT document is single-threaded with other threads in used to handle client requests.
For persistence they can use either an in-memory store, a raw filesystem, or an embedded key-value store.

Additionally, \dismerge{} no longer needs to track the revision counter and related fields: create revision and mod revision for each value.
It also adds the API implementation for tracking replication status with peers as well as logic for calculating the responses.

\subsection{Architecture}

\begin{figure}
    \centering
    \begin{tikzpicture}
        \node[] (client) {};
        \node[draw, rounded corners, fill=orange!50] (capi) [below=1em of client] {Client API};
        \node[draw, rounded corners, fill=orange!50] (kvstore) [below=1em of capi] {KV Store};
        \node[draw, rounded corners, fill=yellow!50] (doc) [below=1em of kvstore] {Document};
        \node[draw, rounded corners, fill=cyan!50] (persister) [below=1em of doc] {Persister};
        \node[draw, rounded corners, fill=orange!50] (watches) [left=of kvstore] {Watches};
        \node[draw, rounded corners, fill=cyan!50] (sync) [right=of kvstore] {Peer sync};
        \node[draw, rounded corners, fill=cyan!50] (papi) [above=1em of sync] {Peer API};
        \node[] (peer) [above=1em of papi] {};

        \draw[-latex] (client) -- (capi);
        \draw[-latex] (peer) -- (papi);
        \draw[-latex] (capi) -- (kvstore);
        \draw[-latex] (kvstore) -- (doc);
        \draw[-latex] (doc) -- (kvstore);
        \draw[-latex] (persister) -- (doc);
        \draw[-latex] (doc) -- (persister);
        \draw[-latex] (doc) -- (watches);
        \draw[-latex] (watches) -- (client);
        \draw[-latex] (papi) -- (sync);
        \draw[-latex] (doc) -- (sync);
        \draw[-latex] (sync) -- (doc);
        \draw[-latex] (kvstore) -- (capi);
    \end{tikzpicture}

    \begin{tikzpicture}
        \node[circle, fill=orange!50] (diffs) {};
        \node[] (diffstext) [right=0.5em of diffs] {Differ between implementations};
        \node[circle, fill=cyan!50] (shared) [below=0.5em of diffs] {};
        \node[] (sharedtext) [right=0.5em of shared] {Shared between implementations};
        \node[circle, fill=yellow!50] (am) [below=0.5em of shared] {};
        \node[] (amtext) [right=0.5em of am] {Automerge};
    \end{tikzpicture}

    \caption{\metcd{} and \dismerge{} architecture.}\label{fig:arch}
    \Description{A graph showing the architecture of the datastores.}
\end{figure}

Figure~\ref{fig:arch} shows the architecture of \metcd{} and \dismerge{}.
Both datastores focus on being \emph{horizontally} scalable rather than \emph{vertically} scalable.
This is in order to span multiple edge sites for availability, rather than large single site deployments.
As such, they do not use up all available cores, instead using only a few threads.
Requests pass through the \etcd{}-compatible gRPC API and into the key-value store.
This key-value store contains an Automerge~\cite{automerge} CRDT document of keys and values.
Changes to the document are prepared in this module before being persisted to disk through the persister.
Once the changes have been persisted they pass back up through the gRPC API to the client.
On the return through the KV store the updated value gets propagated to any watchers and the syncing thread is notified of changes so that it can share the updates with peers.

Operations on the Automerge CRDT document are single-threaded, focusing on limited edge resources, using other available threads to scale client request protocol handling, making changes durable, and communicating with peers.

\subsection{Data model}

Listing~\ref{lst:mergeable-data} shows the data model for \metcd{}, stored in the Automerge document with some example data.
The \verb|kvs| is the main storage for key-value data with each key having a map of the revisions that exist for it.
Deleted values are represented by \verb|null| at the given revision.
This enables efficiently handling queries for current and past data.
Each key can also have an associated lease identifier, which is only applicable to the latest value of the data.
Leases are stored separately in the \verb|leases| key to support efficiently enumerating possible leases in the datastore.
Metadata about the cluster is stored in the \verb|cluster| key including the ID of the cluster and the current revision.
Finally, the list of cluster members is stored in the \verb|members| key, mapping their ID to their name, URLs for peer connections, and URLs for client connections.

\begin{listing}
    \centering
\begin{minted}{json}
{   "kvs": {
        "key1": { "revs": {
                    "001": [118, 97, ...],
                    "003": null },
                  "lease_id": 1 }
    },
    "leases": { "1": null },
    "cluster": { "cluster_id": 2,
                 "revision": 3 },
    "members": {
        0: { "name": "default",
            "peer_urls": [],
            "client_urls": [] }
    }
}
\end{minted}
    \caption{Data model for \metcd{}. Values under \emph{revs} are the encoded bytes.}\label{lst:mergeable-data}
\end{listing}

Listing~\ref{lst:dismerge-data} shows the data model for \dismerge{}, stored in the Automerge document with some example data.
It shares most aspects with \metcd{}'s data model, namely \verb|leases| and \verb|members|.
The \verb|kvs| is the main storage for key-value data with each key storing the latest value and the ID of any lease associated with it, rather than the entire history.
This does not need to store the entire history as that is maintained within and queryable from Automerge directly.
Deleted values have no key in the \verb|kvs| object.
Metadata about the cluster is stored in the \verb|cluster| key but notably no \verb|revision| field is needed compared to \metcd{} as the hashes of the document are obtainable from Automerge.

These data models grow with each client update, enabling historical queries but incurring an overhead to store all the data.
\etcd{} supports compaction of the revision history to reduce the storage space, preventing access to revisions older than the compaction point.
This is not directly supported in \metcd{} or \dismerge{} due to a lack of support for \emph{garbage collection} in Automerge at present, though support is available in other libraries~\cite{yjsgc}.

\begin{listing}
    \centering
\begin{minted}{json}
{   "kvs": {
        "key1": { "value": [118, 97, ...],
                  "lease_id": 1 }
    },
    "leases": { "1": null },
    "cluster": { "cluster_id": 2 },
    "members": {
        0: { "name": "default",
            "peer_urls": [],
            "client_urls": [] }
    }
}
\end{minted}
    \caption{Data model for \dismerge{}. Values under \emph{value} are the encoded bytes.}\label{lst:dismerge-data}
\end{listing}

\paragraph{Consistent initialization}
To ensure that all nodes in a cluster can accept and merge changes from peers they need to start with a consistent state.
Initialization logic on each node sets this up in a consistent way on first start by setting the document's actor ID to 0 and creating empty objects for the key-values, server meta information, members, and leases.
For \metcd{} this initialization also sets the initial revision to 1.
This creates a change with a predictable hash from which all changes can branch off from.

\subsection{API Guarantees}

\begin{table*}
    \centering
    \caption{API guarantee comparison of the datastores.}\label{tab:api-guarantees}
    \begin{tabular}{lllllll}
        \toprule
        Store       & Atomicity & Durability & Consistency     & Write ordering & Watch events          & Revision uniqueness \\
        \midrule
        \etcd{}     & Yes       & Majority   & Linearizability & Total order    & Unordered, complete   & Globally            \\
        \metcd{}    & Yes       & Locally    & Causal          & Partial order  & Unordered, incomplete & Pre-conflict        \\
        \dismerge{} & Yes       & Locally    & Causal          & Partial order  & Unordered, complete   & Globally            \\
        \bottomrule
    \end{tabular}
\end{table*}

While retaining the same wire-level API, the change of consistency model impacts the guarantees that \metcd{} can make.
The adaptations with respect are highlighted in Table~\ref{tab:api-guarantees}.
Atomicity refers to how operations are performed: \metcd{} performs them atomically originally, but merging can make the result non atomic, due to the lack of unique revision addressing.
\dismerge{} provides atomic request handling due to the unique addresses.
For durability, \metcd{} and \dismerge{} both only persist to the local node before returning to the client to avoid reliance on the network connectivity to other nodes.
\metcd{} and \dismerge{} both also provide only partial ordering of writes, that is due to writes being able to be processed at different nodes concurrently, before synchronizing the nodes and merging the data.
Watch events are always unordered, particularly as for \dismerge{} there is no total order to base them off.
Notably, \metcd{} can send incomplete watch events: those that may not contain all of the modifications for that revision related to the watch; this is because merging other changes from peers can mutate an old revision, leading to previously sent watch event being potentially incomplete.
Merging changes in \dismerge{} can never modify an existing revision, and so the watch events are always complete.
Revisions for \metcd{} are also only unique before a node synchronizes with another that has a different operation at the same revision; that is: the revisions are only unique pre-conflict.
\dismerge{} avoids this by bringing back globally unique addresses suitable capturing the causality more accurately.

\subsection{Durability}

\etcd{} stores the contents of the datastore on-disk using the bolt~\cite{bbolt} embedded key-value database.
It uses a flat structure to store the values at all revisions in history, up to the point of the last compaction.
\metcd{} stores values in an Automerge document.
Doing so produces \emph{changes} that encapsulate the operations performed to the document.
It is these \emph{changes} that \metcd{} persists in its embedded key-value database on-disk.
This does mean that the document needs to be loaded into memory before it is queryable, so \metcd{} can end up using more memory than \etcd{} to hold the actual document.
Making CRDTs space-efficient, in both in-memory and on-disk formats, is currently an active area of work~\cite{automerge-binary-format, crdtsthehardparts}.

\subsection{Synchronization}

\begin{figure}
    \includegraphics[]{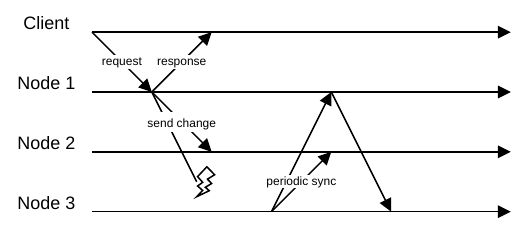}
    \caption{Example of the synchronization process. The message from Node 1 to Node 3 gets lost and later Node 3 obtains the change via periodic sync.}
    \Description{A diagram showing a client sending a request and getting a response. This triggers synchronization from the initial node (node 1) to other nodes (nodes 2 and 3). The message to node 3 gets lost and eventually node 3 performs a periodic sync and obtains it.}
\end{figure}

\begin{figure}
    \includegraphics[width=\linewidth]{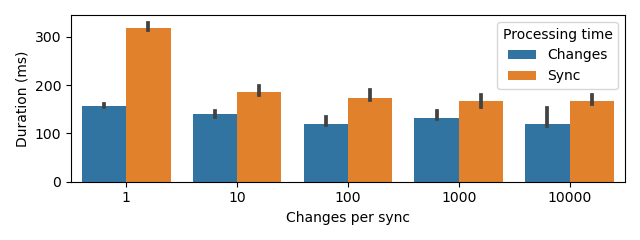}
    \caption{Time spent producing changes and performing periodic synchronization. Two documents concurrently producing an equal number of changes before synchronizing. Each change writes a new value to a shared key. 10,000 changes performed in total with 10 repeats.}\label{fig:sync-latency}
    \Description{A bar plot comparing the number of changes per sync to the duration to perform them. There is a high overhead when using 1 change per sync but this decays with more changes, but remains more costly than the changes.}
\end{figure}

Automerge is an operation-based CRDT, meaning that it only needs to send changes that the peer does not already have, rather than the full state.
\metcd{} and \dismerge{} split synchronization into two main cases: optimistic and pessimistic.
In optimistic synchronization, a node immediately broadcasts a change, generated from a client request, to its synchronization peers.
This enables fast replication in the best-case, when network the network is partition-free.
This method is very simple, making it low-overhead and efficient to implement.
Changes are not forwarded past the initial synchronization peer.
When the network has partitions, these changes may be missed by peers, or peers may not be in the synchronization peers of a node, but should get the change.
To solve this, pessimistic periodic synchronization is performed.
This synchronization uses the protocol built into Automerge, based on Kleppmann and Howard's Byzantine Eventual Consistency protocol~\cite{kleppmann2020byzantine} to synchronize the changes.
The small number of round trips, typically one, required to synchronize aids in minimising the resource requirements and latency when peers have diverged.
Peers propagate all seen changes, enabling transitive connectivity of nodes.
Periodic replication has more overhead than optimistically broadcasting changes as it has to calculate the set of changes to send from the document based on an estimation of what the peer has.
Figure~\ref{fig:sync-latency} highlights this; producing changes is equivalent to the optimistic broadcasting.
Additionally, this has to be done on a peer-by-peer basis, adding extra load with more peer connections.

The topology of a \metcd{} cluster is a complete network.
This is based off of the architecture for \etcd{} since leaders should be able to communicate with a majority of nodes.
However, given \metcd{}'s design to scale horizontally, this communication can quickly become cumbersome due to \(O(n^2)\) connections for \(n\) nodes.
This becomes less of a concern as the synchronization of changes is transitive and the protocol rarely sends changes peers already have.
Alternatively, instead of using a complete network, \metcd{} can be configured with a list of peers to communicate with which form a subgraph of the network.
It is the responsibility of the operator to configure this subgraph and to ensure that there is sufficient redundancy in the deployment.
Future work could extend the peer communication to share addresses of nodes and actively monitor and build a topology based on environmental factors such as latency and redundancy.
This would ease operational aspects of the cluster while also being able to react internally to failures and changes in cluster membership.
However, this is left as future work due to it being highly dependent on deployment scenario.

\subsection{Typing the values}

Treating the values as opaque bytes, as \etcd{} does, can make for efficient handling of requests but forces last-writer-wins semantics when doing conflict resolution with CRDTs.
In practice, these opaque bytes often have a structure similar to JSON, consisting of nested maps and lists.
Since Automerge supports JSON datatypes natively we can offer improved behaviour under conflicting updates to values.
\metcd{} and \dismerge{} clusters can be specialised to custom datatypes for values that will be stored in the cluster.
This specialisation is performed at compile-time using a operator-provided implementation provided in Rust, Listing~\ref{lst:automergeable}.
This implementation is responsible for parsing the bytes from the wire representation into its datatype and updating the stored value in the CRDT, enabling capturing the intent of changes.
For reads, the implementation is responsible for extracting the value from the CRDT and converting it to bytes to send on the wire.
For instance, if updating items in a JSON dictionary then the conflict resolution can allow concurrent edits to different keys easily rather than just accepting one of the objects.
We provide pre-built variants of the datastores supporting raw bytes as well as JSON.\@
Applications using a specialised variant of \metcd{} or \dismerge{} with custom datatypes can also handle translation of data to prior and future schemas as well as validation of datastored.
Using custom datatypes also enables more complex datatypes to be used, for instance using counters rather than plain integers or enriching datastored to support other conflict resolution strategies.

\begin{figure}
    \includegraphics[width=\linewidth]{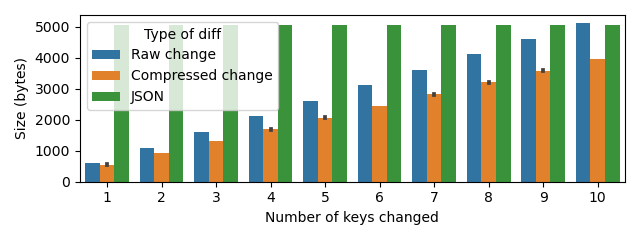}
    \caption{Size of change diff in varying over the number of keys changed. Keys were integers, values were random strings of 500 characters. The JSON case is the size of the total JSON-encoded data.}\label{fig:change-diff}
    \Description{A bar plot comparing the number of keys changed in a request to the size of the operation. When storing entire values it is constant, for Automerge changes it grows linearly with the number of keys modified.}
\end{figure}

Due to the custom datatypes producing minimal \emph{diff}s of the value, this can reduce the amount of data to replicate and persist, Figure~\ref{fig:change-diff} highlights this over a number of keys being changed.
For the edge environment, this can drastically reduce extra traffic between sites, leaving more bandwidth for user traffic.
Each change in the datastore has additional, small, constant overhead beyond the bytes to encode the diff, this is particularly optimised in \metcd{} where multiple client operations are grouped into a single change, whereas each client operation in \dismerge{} creates a new change.

\subsection{Exposed replication status}\label{sec:replication-status}

Now that the datastore's history can be addressed uniquely, we can expose more details to the clients.
One key item is that clients may have differing requirements for the replication of their values before acting on them.
\dismerge{} can accommodate this by informing them of the replication status of a set of frontier hashes.
On each synchronization with peers (periodic synchronization), a node gets an update of what the heads of the other nodes are, this also includes a notion of what frontier hashes both nodes have in common.
From this, and a set of frontier hashes a client is interested in, the node can calculate which peer nodes have the change.
This is limited to direct peers of a node but clients can iteratively query other nodes to gather more information if desired.
With this information, clients can dynamically choose their replication factor without placing a significant extra burden on the server.
This API is available as a unary endpoint where the client sends a request for a set of frontier hashes and receives a single response indicating, for each peer, whether they have the change corresponding to the hash.

\subsection{Model overheads}\label{sec:model-overheads}

\begin{figure}
    \includegraphics[width=\columnwidth]{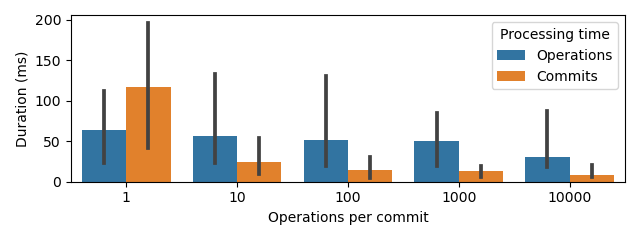}
    \caption{Time spent on operations and commits in Automerge varying operation counts per commit. Each operation writes to the same key in a \emph{map}. Run for 10,000 operations with 100 repeats.}\label{fig:commits}
    \Description{A bar plot comparing the number of operations per change to the time spent on them. In particular it compares time spent on operations vs the final commit. For 1 operation per commit there is a big overhead: commits cost double the time for just the operations. This quickly decreases to become negligible.}
\end{figure}

Since \metcd{} does not leverage the hash graph of Automerge it can batch multiple operations into a single \emph{change}.
By leveraging the hash graph for addressing \emph{changes}, \dismerge{} requires each client operation to be in a separate \emph{change}.
This leads to a trade-off in the time spent processing the operations and the overhead of \emph{committing} each \emph{change}, explored in Figure~\ref{fig:commits}.
During committing of a change there is a need to calculate the hash of the encoded representation.
This adds an overhead to processing a given number of client requests to serialize metadata for the change as well as the operation, before hashing.
This can also impact the performance of individual operations due to the cache locality of data.
Lower level changes in Automerge may be possible to optimise the overhead of calculating the hash but we do not delve into this in this work.

\section{Evaluation}~\label{sec:eval}

We evaluate both \metcd{} and \dismerge{} in comparison to \etcd{} starting at an edge-like deployment and then working towards a single node setup.

\begin{enumerate}
    \item How do \metcd{} and \dismerge{} handle a partition compared to \etcd{}, particularly at scale? Section~\ref{sec:clustered-partition}
    \item Assuming a reliable network without partitions, how does this change the performance of \etcd{} at scale compared to the others? Section~\ref{sec:clustered-delay}
    \item How would this performance differ if we were in a data-center-like environment? Section~\ref{sec:clustered}
    \item What overhead do \metcd{} and \dismerge{} add for single-node performance, given that clients will be working with their local node? Section~\ref{sec:single-node}

\end{enumerate}

\subsection{Setup}\label{sec:eval-setup}

Benchmarks were run on a single Azure \verb|Standard D64ds v5| (64 vcpus, 256 GiB memory) machine, running Ubuntu 20.04, with 3 repeats.
Load is generated using an open-loop load generator and uses the YCSB workload A, which issues an equal ratio of updates and reads uniformly spread across the keyspace.
All requests were sent to a single node, to mimic a workload at a single edge site, and load was sustained for 5 seconds.
Keys are 18 bytes and values are 32 bytes, randomly generated.
Each datastore node was run in a Docker container and limited to 2 CPUs to mimic limited edge resources.
The datastore nodes are backed with a \emph{tmpfs} to minimise the impact of disk latency.
No additional latency is added between the nodes unless specified.
All results presented are for successful requests.
The setup models a client interacting with its local datastore node only, relying on it to process the operations.
The client initially connects directly to the local leader node, this avoids forwarding overhead in \etcd{}.
When the leader node is partitioned from the rest of the cluster, the leader will change and, after the partition heals, the client may be connected to a non-leader node.
Partitions were injected with the use of \verb|iptables|, delays were injected with the Linux \verb|traffic controller| with a variation of 10\% and a correlation of 25\%.

\subsection{Starting at the edge}\label{sec:clustered-partition}

At the edge, applications will be deployed across sites, needing to share data between these.
The sites are geographically distributed with limited resources at each, network links can also be unreliable.
As such, this section works within this context with a setup of three nodes spread over sites, connected over a 10ms link.
The client is co-located with a node, initially the leader node and a partition is injected between the leader node and the rest of the cluster at approximately 5 seconds, before being healed at 10 seconds into the experiment.

\begin{figure}
    \begin{subfigure}[b]{\linewidth}
        \includegraphics[width=\linewidth]{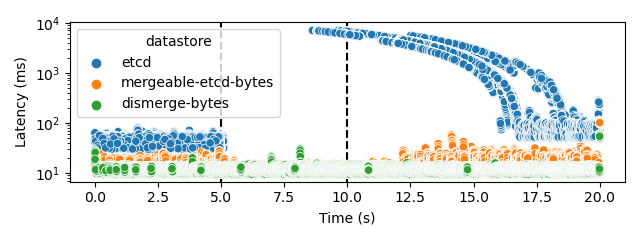}
        \caption{Latency of successful requests.}\label{fig:clustered-delayed}
        \Description{A scatter plot showing etcd is unable to handle requests during a partition.}
    \end{subfigure}
    \begin{subfigure}[b]{\linewidth}
        \includegraphics[width=\linewidth]{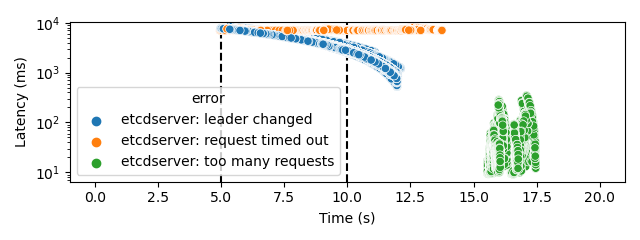}
        \caption{Latency of failed requests by error condition, only from \etcd{}.}\label{fig:clustered-delayed-errors}
        \Description{A scatter plot of showing the etcd errors that occur from the partition and then from the overload after partition recovery.}
    \end{subfigure}
    \caption{Workload applied to a three node cluster. The leader node is partitioned from the cluster at approximately 5 seconds into the experiment, and this is cleared at 10 seconds in (dashed vertical lines). The \(y\) axis is log-based.}
\end{figure}

Figure~\ref{fig:clustered-delayed} shows the results of this experiment for each datastore.
Initially, \etcd{} has a higher latency due to the latency of the network between the nodes.
During the partitioned period \etcd{} is unable to service requests, internally queueing them until they time out.
This is what leads to some requests issued before the partition heals to be processed.
When the partition is healed the local node also has an overload of requests, as shown by the \enquote{too many requests} errors in Figure~\ref{fig:clustered-delayed-errors}.
During this recovery time, the local node is also trying to obtain who the new leader is and forward requests to them for processing.
This further exacerbates the latency of successful requests, and leads to more overload.
Requests that end up being successfully handled after the partition is healed and a steady state is obtained now incur higher latency as the local node is no longer a leader, it must forward each request.

\metcd{} and \dismerge{} are able to continue processing requests during the partition, holding changes to be synchronized until the partition heals.
This maintains reliable performance during the disruption and avoids costly recovery overheads after.
The periodic synchronization will ensure that replicas obtain all of the missed changes.

\subsection{Making the network reliable}\label{sec:clustered-delay}

\begin{figure}
    \includegraphics[width=\linewidth]{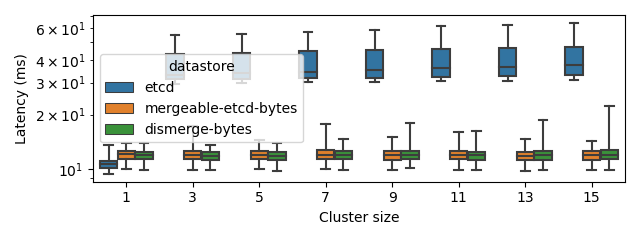}
    \caption{Latency box plot of multiple nodes with 10ms latency on each link. Whiskers extend from the 1st to the 99th percentile. The \(y\) axis is log-based.}\label{fig:multi-node-delayed}
    \Description{A box plot showing etcd's increased latency passed on from the network.}
\end{figure}

Assuming that the network will be reliable, not experiencing partitions, we can view how the latency of the network affects the scale of the cluster more directly.
This setup follows that of the previous section but no partition is injected during the experiment run, and so the leader node remains stable.
Due to \etcd{}'s eager replication, it is very sensitive to the performance of the network.
Figure~\ref{fig:multi-node-delayed} presents plots of the latency distribution and peak throughput across different cluster sizes.
Cluster sizes are generated from the \(2f+1\) function for \etcd{} to maximise failure tolerance for \(f\) failures.

For single node deployments there is no network latency incurred as no replication is performed.
However, when adding nodes \etcd{}'s latency drastically increases due to its requirement to replicate data to a majority of nodes in the processing of a request.
As the cluster size increases, this incurs a marginal overhead to communicate with the nodes in the cluster.
This highlights \etcd{}'s sensitivity to the network latency for processing requests.
This also makes the assumption that all links are homogeneous, in reality they are likely to be heterogeneous due to their geographical distribution and so some remote nodes could drastically impact the latency characteristics.
This is further worsened when the leader changes as it could change to a site with slower connections to a majority, bottlenecking all requests on a single slow link.

\metcd{} and \dismerge{}, moving eager communication off the critical path, enable more consistently low-latency operation, even at larger cluster scales.
They too will incur an overhead of communicating with a larger number of peers but this is expected to be significantly lower than the delay added to \etcd{} due to the network latency.
This can also be managed by not connecting all nodes to all nodes, instead forming a mesh network.

\subsection{Providing an optimal network}\label{sec:clustered}

\begin{figure}
    \includegraphics[width=\linewidth]{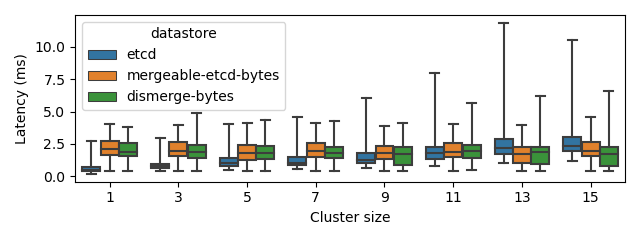}
    \caption{Latency box plot of multiple nodes. Whiskers extend from the 1st to the 99th percentile.}\label{fig:multi-node}
    \Description{A box plot showing an increase in etcd's latency with increasing cluster scales even without network latency.}
\end{figure}

Since \etcd{} is targeted for cloud data center deployments we now evaluate its scalability in a setting with no latency, but still limited resources.
This also highlights the overhead of added fault tolerance, something which may still be important to cloud applications and which may limit the resources each node can have.
The impact of varying the cluster sizes can be observed in Figure~\ref{fig:multi-node}, under a target rate of 10,000 requests per second.
Generally, \etcd{} encounters scaling issues in terms of latency with the increase in cluster size.
Due to \etcd{}'s optimised implementation, \metcd{} and \dismerge{} currently have a higher, but still small, fixed cost.
Despite this and our analysis in the previous section suggesting that the overhead of communicating with more nodes is marginal for \etcd{}, we observe that there is indeed an overhead incurred by \etcd{} which seems to be non-trivial compared to the performance of small clusters.
This trend implies a cross-over point where clusters of \etcd{} with no latency overhead become less performant than \metcd{} and \dismerge{}.
We project \etcd{}'s latency to continue to get worse as cluster size increases due to the fundamentally increasing amount of work that the leader node must perform to replicate values and the eager nature of this.

\subsection{Collapsing the cluster}\label{sec:single-node}

\begin{figure}
    \begin{subfigure}[b]{\linewidth}
        \includegraphics[width=\linewidth]{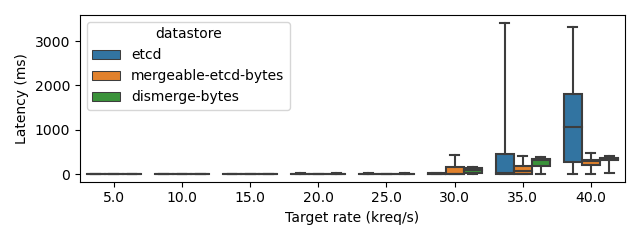}
        \caption{Latency box plot. Whiskers extend from the 1st to the 99th percentile.}\label{fig:single-node-latency}
        \Description{A box plot showing latency at high throughputs. All datastores increase in latency from around 30,000 requests per second.}
    \end{subfigure}
    \begin{subfigure}[b]{\linewidth}
        \includegraphics[width=\linewidth]{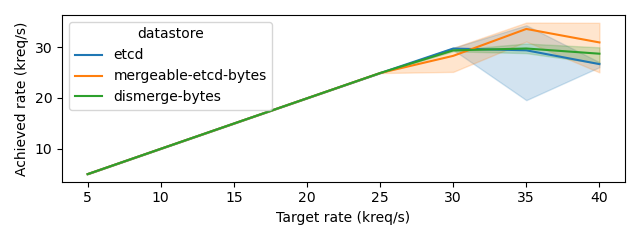}
        \caption{Comparison of achieved rate with respect to the target rate. Repeat variance shown by the shaded region.}\label{fig:single-node-throughput}
        \Description{A line plot showing the achieved rate of the datastores.}
    \end{subfigure}
    \begin{subfigure}[b]{\linewidth}
        \includegraphics[width=\linewidth]{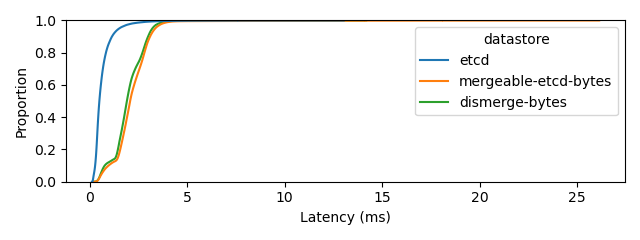}
        \caption{Latency CDF at 10,000 requests per second to highlight differences at lower loads.}\label{fig:single-node-cdf}
        \Description{A latency CDF showing the slightly higher latency of mergeable-etcd and dismerge.}
    \end{subfigure}
    \caption{Single node results.}\label{fig:single-node}
\end{figure}

To compare the raw overhead of the data model that \metcd{} and \dismerge{} use internally we compare results of a single node handling requests.
This avoids conflation with the synchronization process.
From Figures~\ref{fig:single-node-latency} and~\ref{fig:single-node-throughput}, we observe that all datastores can handle the load up to around 30,000 requests per second, after which throughput drops off for all.
However, after this point \etcd{} suffers significantly higher latency, not efficiently shedding or rejecting load.
We can also observe the higher overhead within \dismerge{} compared to \metcd{} at higher rates due to the overhead of extra commits, discussed previously in Section~\ref{sec:model-overheads}.

Looking at Figure~\ref{fig:single-node-cdf} we can observe that for lower rates \etcd{} outperforms both \metcd{} and \dismerge{} in terms of latency.
This is expected due to the extra overheads that the CRDT logic impose upon \metcd{} and \dismerge{}.
When processing a write request, \etcd{} simply needs to write it to the in-memory maps and caches before persisting the write, which is effectively a no-op due to the \emph{tmpfs}.

Errors begun to occur from the datastores from 30,000 requests per second.

\section{Implications on applications}\label{sec:discussion}

Given the trade of linearizability for causal consistency applications may not function correctly without modification.
More subtly, the difference in replication of data from eager to lazy can impact the durability guarantees of applications.
Taking \kubernetes{} as an example application that already relies on \etcd{}, we can imagine replacing it with \metcd{}.
\kubernetes{} primarily operates by storing a desired state of resources in \etcd{}.
This central view of the data is then acted upon by \emph{controllers} that \emph{watch} the data and react to new versions.
These controllers perform operations such as creating new container resources (\emph{Pods}) which are delivered to the scheduler and then allocated to a node.
\emph{Controllers} also handle higher-level resources such as \emph{Deployments} which dictate the number of containers running in the cluster for a particular application.

Due to the controllers present in \kubernetes{}, we believe that \metcd{} would enable it to remain functional.
Any discrepancy of the data due to merging of concurrent interactions will be acted upon by the \emph{controllers} and corrected.
Importantly, \metcd{} does not impact the integrity of the values, only which value would be presented.
It is unclear whether \kubernetes{} provides stable and sufficiently dampened control loops to handle higher latency between the datastore nodes, and thus more temporary divergence.

Under this model every partition of the datastore cluster effectively creates a replica of the entire cluster, starting new instances of applications on both sides of the partition to ensure replica counts are met.
When the partition heals and the datastore nodes synchronize, there will be one cluster again the controllers will drive the state to that of the single cluster again.

One problematic piece of \kubernetes{} would be its guarantee of unique \emph{Pod} names.
This is typically not an issue as \emph{Deployments} create \emph{Pods} with randomised names, preventing collisions.
However, \kubernetes{} manages stateful deployments with a \emph{StatefulSet} which assigns numerically increasing names for the \emph{Pods}.
This could lead to multiple \emph{Pods} with the same name existing in the cluster due to the weaker consistency in the datastore.
One possible mitigation is to have the site-local \emph{StatefulSet} controllers only manage the instances at their site, injecting a suffix for the site name into the pod name to make them unique again.

\kubernetes{}, storing resource definitions as a JSON-like protobuf schema, would be a prime candidate for exploring the use of the typed values in \metcd{}.
For instance, replica counts on \emph{Deployment} resources could be modified concurrently to the other fields, such as the container image to be run.
This enables concurrent updates to take effect, rather than requiring the initiators to retry their requests.
For \emph{Deployments} this is of interest to even higher-level controllers that might be in charge of updating the image or providing dynamic scaling.

Integrating \dismerge{} into \kubernetes{} would be more invasive due changes in how history is addressed, but should be feasible under the above discussion and ultimately lead to a more intuitive model due to its immutable history.

\section{Related work}\label{sec:related}

Anna~\cite{annakvs} is a distributed key-value store that targets performance at both single node and cloud-scale through a system of coordination-free actors.
Anna also uses CRDTs for storage though uses a custom implementation rather than a library.
Anna focuses on the core functionality of a distributed key-value store, not implementing related functionality such as watching keys.
As such, it is not a direct competitor to \metcd{} but provides good lessons if \metcd{} were to need scaling to cloud-scale workloads.

Azure's CosmosDB~\cite{cosmosdb} is a closed-source NoSQL database that provides many different consistency levels and with different API compatibility layers.
This allowed CosmosDB to expose an \etcd{}-compatible API whilst changing the consistency levels dynamically~\cite{cosmosdbetcd}.
The database can also produce reports of the staleness of the data returned, enabling insight into the support of the application for weaker consistency levels which may lead to performance improvements.

\section{Conclusion}\label{sec:conclusion}

Deploying platforms and applications near to the edge provides new challenges in delivering higher-level requirements.
From these we derive lower-level key-value datastore requirements and show how \etcd{} is unsuited to meet these.
We explore the design space under these requirements, focusing on consistency, history addressing, durability and value representation.
This exploration then instigates the implementation of two new datastores, successively adapting \etcd{} to be edge-suitable: \metcd{} and \dismerge{}.
These datastores offer applications reliable local-first operation, enabling applications to continue operating under unreliable network conditions found at the edge.
The performance is also considerably enhanced compared to \etcd{}, providing consistent low-latency operation.
Due to \etcd{}'s popularity as a critical distributed key-value store, we envision new avenues for work focusing on local-first edge applications, avoiding eager coordination with other sites.
Furthermore, this can be extended to cloud environments to enhance reliability as both \metcd{} and \dismerge{} offer competitive performance and can be scaled to much greater extents.
More broadly, this work highlights a transition from servers being co-located with each other with distributed clients, to servers being co-located with clients but being distributed from other servers.

\clearpage
\bibliographystyle{plain}
\bibliography{default.bib}

\end{document}